\documentclass[aps,prl,graphicx,amsmath,amssymb,reprint]{revtex4-2} % Two-column 'reprint'

\usepackage{graphicx}
\usepackage{subfig}
\usepackage{mathtools}

\usepackage{dcolumn}% Align table columns on decimal point
\newcolumntype{d}[1]{D{.}{\cdot}{#1} }
\usepackage{bm}% bold math
\usepackage{xcolor,soul}

\usepackage{siunitx}% JMF addition - once a physicist, always a physicist
\begin{document}
\title{Anharmonic phonons with Gaussian processes}
\date{\today} % useful to track hard copy of drafts

\author{Keerati Keeratikarn}
\affiliation{Department of Physics, Imperial College London, Exhibition Road, London  SW7 2AZ, UK}

\author{Jarvist Moore Frost}
\affiliation{Department of Chemistry, Imperial College London, UK}
\altaffiliation{Department of Physics, Imperial College London, Exhibition Road, London  SW7 2AZ, UK}

\email[Electronic mail:]{jarvist.frost@imperial.ac.uk}

\keywords{Potential Energy Surface, Anharmonic Force Constants, Gaussian Processes}

\begin{abstract}

We provide a method for calculating anharmonic lattice dynamics, by building a surrogate model based on Gaussian Processes (GPs). Due to the underlying Gaussian form of a GP, the model is infinitely differentiable. This allows us to train the model trained directly on forces (the derivative of PESs) reducing the evaluations required for a given accuracy. We can extend this differentiation to directly calculate second and third order force-constants using automatic differentiation (AD). For the five model materials we study, we find that the force-constants are in close agreement with a standard finite-displacement approach. Our method appears to be linear scaling in the number of atoms at predicting both second and third-order (anharmonic) force-constants.
%and of cluster-expansion approach (in \texttt{HiPhive}). 

\end{abstract}

%\pacs{71.38.-k, 71.20.Nr, 71.38.Fp, 63.20.Kr}
% 71.38.-k	Polarons and electron-phonon interactions
% 71.20.Nr 	Semiconductor compounds 
% 71.38.Fp	Large or Fröhlich polarons
% 63.20.Kr	Phonon-electron and phonon-phonon interactions

% 72.40.+w 	Photoconduction and photovoltaic effects
% 61.66.Fn 	Inorganic compounds 
% 88.40.jn 	Thin film Cu-based I-III-VI2 solar cells
% 88.40.-j 	Solar energy

\keywords{Anharmonicity; Gaussian processes; Phonon}

\maketitle 

% ? is Process in Gaussian Process capitalised

\section{Introduction}\label{introduction}

Finite temperature properties of matter require a model for thermal motion.
In crystals it is natural to describe the thermal motion as collective excitation around the equilibrium structure.
The standard approach is to use a finite displacement method (FDM), based in a Taylor expansion of the potential energy surface around the equilibrium. 
The second-order force constants provide the harmonic approximation for the potential energy surface (PES); 
anharmonic contributions (required for finite thermal conductivity) require higher order force constants.
FDM-based calculations scale poorly both with the size of the system, and the order of the force constants.

The potential energy surface (PES) of a solid-state material directly provides the mechanical response of a material (Young's modulus, yield strength), and also provides the environment in which the electronic structure exists. 
Thus it is key to both the electron-phonon coupling which lead to finite electrical conductivity, and the phonon-phonon scattering which lead to finite thermal conductivity. 

For small perturbations around the equilibrium structure, a Taylor expansion of the potential energy surface can be made. 
As at equilibrium the first-order (linear) term of this expansion is zero (i.e. no forces), the first non-zero term is the second-order harmonic contribution. 
Cutting off the expansion at this point (the harmonic approximation) leads to the mass-weighted `dynamic matrix' of force-constants. 
Diagonalising this matrix produces normal-modes (eigenvectors of the dynamic matrix) and vibration frequencies (eigenvalues of the dynamic matrix), which describe the phonon properties of a material. 
This change of basis is often used as a natural way in which to describe the response theory of solid state materials. 
The phonon modes can be directly measured as a function of crystallographic momentum with neutron scattering experiments, and the gamma-point modes give rise to infrared absorption of materials (by the polarisation of a given mode), and Raman response (by the hyperpolarisation of a given mode).

These normal-modes do not interact, and therefore have an infinite lifetime; thus a harmonic picture of matter directly leads to the prediction of infinite thermal conductivity. 
Anharmonic corrections to this picture can be added via many-body perturbation theory, where second quantisation is used to consider these vibration states as quasi-particles, the phonon force Boson. 

The standard approach to calculate phonons is via a finite displacement method. 
For harmonic phonons, this requires moving each atom a small displacement separately in ${x,y,z}$, requiring $3^2N^2$ evaluations of the total energy. 
(Typically these forces come from an electronic structure method, such as density functional theory.) 

To include third-order force-constants in a finite displacement method, one must consider each atom moving in ${x,y,z}$ combined with every other atom moving in ${x,y,z}$, requiring $3^3N^3$ separate evaluations of the total energy. 
% JMF: PK, can you check this maths? I think it's right

This prohibitive scaling, combined with the minimal $O(N^3)$ scaling of electronic structure methods, severely limits the size of systems for which we can predict the thermal conductivity. 
This is a major limitation in humanity's ability to design materials with specific thermal characteristics. 
A particular technical application is in the design of thermoelectrics, where one wants to minimise thermal conductivity while maximising electrical conductivity, and for which state-of-the-art materials are extremely complex to maximise phonon scattering.

From an information theoretic point of view, this restriction is curious. 
There is no more information embedded in the third-order derivatives than the second. 
In fact, the full potential energy surface is present in every calculation, and in electronic structure methods the evaluation of the more complicated system has already been paid for by the $O(N^3)$ scaling of the electronic structure method. 

As an alternative approach, one could use a more sophisticated surrogate method, than fitting the individual terms of a Taylor expansion. 
In this work we use Gaussian Processes as a machine-learning surrogate potential energy surface. 
Gaussian processes (GPs) are a Bayesian (probabilistic) machine learning method which can fit an arbitrary function.
During the last 15 years, a number of methods have been developed using Gaussian Processes to fit potential energy surfaces\cite{Handley2009, Bartk2010, Bartk2013, Bartk2015, Cui2016, Dral2017, Guan2017, Kolb2017, Guan2018, Bartk2018, Strickson2019, Dai2020, Sugisawa2020}.

Our particular focus is motivated by the fact that every linear operation on a Gaussian process results in a (transformed) Gaussian process. 
Differentiation is one such linear operation. 
In the context of force-constants, this means that the hierarchy of force-constants ($\frac{\partial^N E}{\partial r^N}$) can simultaneously be used to condition (fit) the Gaussian Process energy surface; and to be directly calculated from the Gaussian process. 
This enables a maximal amount of information to be extracted from each individual electronic structure calculation; and for the calculation of force-constants (of arbitrary order), without further calculation.
As a particular technical application, this may enable the calculation of anharmonic phonons for larger unit cells at a lower computational cost. 

%\red{JMF: Text below needs to be made more specific, and more descriptive as to what has been done before, and to separate it from what we do here.}

Prior work in this area includes Gaussian Processes have recently been applied to fitting the low-dimensional potential energy surfaces for polarised water \cite{Handley2009}, $\mathbf{N}_{4}$ \cite{Cui2016}, a single Si-crystal \cite{Bartk2018,Strickson2019}, methyl chloride \cite{Dral2017} and even chemical reactions \cite{Kolb2017, Guan2017, Guan2018}. 
Wu, Aoi and Pillow provide a general Bayesian optimisation method that makes use of gradient information \cite{1704.00060}. 
This follows from earlier work which concentrated on low-dimensional observations of a dynamic model \cite{Solak2002}. 
Gaussian process models of potential energy surfaces and forces have found considerable application in molecular dynamics of materials \cite{Bartk2010, Bartk2013}.

Gaussian processes scale poorly with respect to the number of training points $m$, as $O(m^{3})$ from an inversion of an $m\times m$ matrix to normalise the model and $O(m)$ from a dot-product of $1\times m$ matrix to evaluate the model \cite{Sugisawa2020, RasmussenW06}.

Several techniques are used to minimise the amount of training data required.
One such technique is to increase the complexity of Gaussian processes model by compositing two or more kernel functions together.
Dai and Krems \cite{Dai2020} illustrated the use of composite kernel to interpolate and extrapolate the 6-dimensional potential energy surfaces of $\mathbf{H}_{3}\mathbf{O}^{+}$.
Subsequently, a 51-dimensional potential energy surface for protonated imidazole dimer is predicted via the full-dimensional GP model compositing several GP models contributing to lower-dimensional molecular fragments, presented by Sugisawa et al. \cite{Sugisawa2020}.

%Gaussian Processes have recently been applied to fitting the potential energy surface of poly-atomic molecules\cite{Dai2020}. % TODO: read this & see whether they differentiated the GP
%Sugisawa et al.\cite{Sugisawa2020} build a Gaussian Process model for a 51 dimension energy surface.
% TODO - read this paper and see what they did :^)

Differentiating the Gaussian Process is another approach to reduce the training data required. 
Garijo del R{\'{\i}}o et al.\cite{GarijodelRo2019} used Gaussian processes as a surrogate model of the anharmonic potential energy surface to accelerate structural relaxation, providing the \textsc{GPMin} function in the Atomic Simulation Environment (ASE) package. 
However, their Gaussian process directly fits a $1+3N$ vector of the total energy and the $3N$ forces. 
They do not exploit the linear nature of the differentiation operation, and therefore the ability to build a nested Gaussian process.

More similar to our work, 
Kaappa et al.\cite{Kaappa2021} used a gradient-aided model to enhanced global structure prediction.  
Asnaashari and Krems\cite{Asnaashari2021} added composite kernels to boost the accuracy of force and energy prediction of large molecules. 

As far as we know, we are the first people to take this approach to higher derivatives (anharmonic force constants), and apply them to `lattice dynamics'.

% TODO: add discussion of compressed sensing at this point? 

%\red{JMF: Would be good to have a quick discussion of the literature in compressed sensing / HiPhive approach, possibly also the sampling from molecular dynamics (Togo, Dynaphonopy) method.}

%\red{Still seemed a bit confused? What of Compressed sensing?}

An alternative approach to accelerate anharmonic calculations is that of the \textsc{HiPhive} \cite{Eriksson2019} extension of cluster-expansion.
The potential of the atomic environment is constructed from a sum of two-body, three-body and four-body clusters.
In these clusters, the pair (or the group) of atoms are contributed by harmonic (or higher anharmonic) force constants. 
These force constants are parameterised by training machine learning on prior forces associating to the input configuration in the atomic environment.

\section{Methods}

The key development challenge in our work was in building derivatives into the Gaussian Processes: this we did by a combination of analytic derivatives of the Gaussian Process kernels, and by using Automatic Differentiation (which automatically applies the chain rule of derivation recursively) at the computer code level. 
McHutchon's technical note was extremely useful in providing a number of Gaussian Process kernel derivatives\cite{mchutchon_dierentiating_2013}.

\subsection{Gaussian Process model}
We begin with discussing standard Gaussian process regression, in the context of the potential energy surface and its derivatives.

To predict potential energy surfaces by Gaussian processes, a potential energy $E(\mathbf{x}^{*})$ can be written in terms of the sum of arbitrary basis functions $\phi_{h}$ as a function of an atomic configuration (descriptor) $\mathbf{x}^{*}$, which can be expressed as  
\begin{equation}
    \label{eqn:1}
    \begin{split}
        E(\mathbf{x}^{*})&=\sum_{h\:=\:1}^{H}w_{h}\phi_{h}(\textbf{x}^{*})\\
                &=\textbf{w}^{\text{T}}\Phi(\textbf{x}^{*}).
    \end{split}
\end{equation}
The descriptor $\mathbf{x}^{*}$ can be chosen to represent a set of atomic positions in Cartesian coordinates for an $N$-atom system, 
\begin{equation}
    \label{eqn:2}
    \begin{split}
        \mathbf{x}^{*} := \{x^{*}_1, y^{*}_1, z^{*}_1, ... ,x^{*}_N, y^{*}_N,z^{*}_N\} .
    \end{split}
\end{equation}
The parameters $w_{h}$ of an $H$-dimensional weight vector $\mathbf{w}$, corresponding to a basis vector $\Phi$, can be selected to be a Gaussian (normal) distribution weight of prior potential energy surface data $E(\mathbf{x})$.
Each Gaussian weight is distributed as $P(w_{h}):=\mathcal{N}(w_{h};\:0,\:\sigma_{w})$ which has a zero-mean and an individual variance $\sigma_{w}^{2}$. 
Regarding the expression in (\ref{eqn:1}), we can write the covariance between a prior potential energy $E(\mathbf{x})$ and a posterior $E(\mathbf{x}^{*})$ as
\begin{equation}
    \label{eqn:3}
    \begin{split}
        \left\langle E(\mathbf{x})E(\mathbf{x}^{*})\right\rangle &=\sum_{hh'}^{H}\langle w_{h}w_{h'}\rangle \phi_{h}(\textbf{x})\phi_{h'}(\textbf{x}^{*})\\
        &=\sum_{hh'}^{H} \left(\int \text{d}^{H}\mathbf{w} P(\mathbf{w})w_{h}w_{h'}\right)\phi_{h}(\textbf{x})\phi_{h'}(\textbf{x}^{*}) .
    \end{split}
\end{equation}
Since the covariance between two Gaussian weights is to integrate them over the distribution, the integral yields $\sigma_{w}^{2}\delta_{hh'}$ so we have 
\begin{equation}
    \label{eqn:4}
    \begin{split}
        \left\langle E(\mathbf{x}),\:E(\mathbf{x}^{*})\right\rangle 
        &=\sum_{hh'}^{H} \left(\sigma_{w}^{2}\delta_{hh'}\right) \phi_{h}(\textbf{x})\phi_{h'}(\textbf{x}^{*})\\
        &=\sigma_{w}^{2}\sum_{h}^{H} \phi_{h}(\textbf{x})\phi_{h}(\textbf{x}^{*}).
    \end{split}
\end{equation}
The inner product between two basis functions is defined as the similarity measures of two atomic descriptors and be rewritten as a kernel (covariance) function
\begin{equation}
    \label{eqn:5}
    \begin{split}
        k(\mathbf{x},\:\mathbf{x}^{*})=\sum_{h\:=\:1}^{H}\phi_{h}(\textbf{x})\phi_{h}(\textbf{x}^{*}).
    \end{split}
\end{equation}

We can use this kernel function to perform Gaussian process regression for predicting a posterior potential energy $E(\mathbf{x}^{*})$ from a prior dataset of previous calculations of potential energy surfaces $\mathbf{E}$. 
The prior dataset consists of $m$ potential energy surfaces corresponding to $m$ atomic descriptors, i.e. $\mathbf{E} := \left[E_{1}\:E_{2}\cdots E_{m} \right]^{\text{T}}$ and $\textbf{X}:= \{\textbf{x}_{1},\:\textbf{x}_{2},\cdots,\:\textbf{x}_{m}\}$ respectively.
Each prior $E_{i}$ can have an observation noise $\sigma_{e}$, due to the error or convergence limitation of the electronic structure calculations: $E_{i} =E(\mathbf{x}_{i})+\sigma_{e}$.
If the prior $E_{i}$ is (\textit{univariate}) Gaussian distributed independently for each data, we will obtain the covariance of any two of the prior data as $\langle E_{i}E_{j}\rangle = k(\mathbf{x}_{i},\:\mathbf{x}_{j})+\sigma^{2}_{e}\delta_{ij}$ 
and can write the prior probability distribution of each $E_{i}$ with zero mean and variance $\sigma^{2}_{e}$ as 
\begin{equation}
    \label{6}
        P(E_{i}) = 
        \frac
        {\textbf{e}^{
        \left({-\frac{1}{2}
        (k(\mathbf{x}_{i},\:\mathbf{x}_{i})+\sigma^{2}_{e})^{-1}
        \cdot E_{i}^2}
        \right)}}
        {\sqrt{(2\pi)\:(k(\mathbf{x}_{i},\:\mathbf{x}_{i})+\sigma^{2}_{e})}} .
\end{equation}

Therefore, the probability distribution of all priors (\textit{marginal likelihood}) is a \textit{multivariate} Gaussian distribution with zero mean and the $m\times m$ covariance matrix, $\langle \textbf{E}\textbf{E}^{\text{T}}\rangle = K(\textbf{X},\textbf{X})+\sigma^2_{e}\mathbb{I}_{mm}$, expressed as
\begin{equation}
    \label{7}
        P(\textbf{E}) =
        \frac
        {\textbf{e}^{
        \left(-\frac{1}{2}
        \textbf{E}^{\text{T}}
        \cdot
        \left[K(\textbf{X},\:\textbf{X})+\sigma^2_{e}\mathbb{I}_{mm}\right]^{-1}
        \cdot
        \textbf{E}
        \right)}}
        {\sqrt{(2\pi)^{m}\:\text{det}\left[K(\textbf{X},\:\textbf{X})+\sigma^2_{e}\mathbb{I}_{mm}\right]}}.
\end{equation}

To make a Gaussian process regression function $E(\mathbf{x}^{*})$, we use the analogy of \textit{Bayesian inference} to calculate the conditional probability distribution of $E(\mathbf{x}^{*})$, given by the prior $\mathbf{E}$:
\begin{equation}
    \label{8}
     \begin{split}
        P\left(E(\mathbf{x}^{*})|\textbf{E}\right)
        = \frac{P\left(\textbf{E},\:E(\mathbf{x}^{*})\right)}{P\left(\textbf{E}\right)} .
    \end{split}
\end{equation}

This is also normal distributed since $P\left(\textbf{E},\:E(\mathbf{x}^{*})\right)$ is a Gaussian joint distribution.
By substituting (\ref{7}) into (\ref{8}) and manipulating it algebraically following the derivation in \cite{MackayD}, we can write the posterior mean of $\left(E(\mathbf{x}^{*})|\textbf{E}\right)$ as
\begin{equation}
    \label{9}
        E(\mathbf{x}^{*}) = k(\textbf{x}^{*},\:\textbf{X})\cdot \left[K(\textbf{X},\:\textbf{X})+\sigma^2_{e}\mathbb{I}_{mm}\right]^{-1}\cdot\textbf{E},
\end{equation}
where $k(\mathbf{x}^{*},\;\textbf{X})$ is a $1\times m$ covariance matrix between the posterior $E(\mathbf{x}^{*})$ and the dataset $\mathbf{E}$.

Gaussian processes are obviously powerful for predicting potential energy surfaces, since the model depends only on a kernel as function of two (same and different) atomic descriptors and is trained on the prior potential energy surface data directly.
There is no need to determine the basis functions and the corresponding weights explicitly for the prediction. 

However, it is possible to define a basis set in order to model nonlinear functions where there is nonlinear relationship among the corresponding descriptors. 
For instance, instead of representing atomic configurations in the Cartesian basis we could use \textit{spherical harmonics} as a basis function, making the model equivariant, as used to great effect in recent works orientated towards molecular dynamics\cite{Bartk2013,Drautz2019,MACE,MACE2}.
This approach is also known as the ``\textit{kernel trick}''.
%This, additionally, means the physical interpretation and understanding may not be required to predict a potential energy.

Commonly used Gaussian Process kernel functions allow us to define the similarity measure between an atomic descriptor in our Gaussian process model. 
Essentially these contain a prior belief in how much influence nearby data points should have on the underlying model. 
They generally encode the belief that physics is myopic---nearby data-points should have a greater influence on the model. 

The standard and mathematically convenient kernel is the \textit{squared exponential} (\textit{Gaussian or Radial Basis Function: RBF}) kernel,
\begin{equation}
    \label{10}
        k(\mathbf{x},\:\mathbf{x}^{*})=\sigma_{o}\exp{
        \left(-\frac{|\mathbf{x}-\mathbf{x}^{*}|^{2}}{2l^{2}}\right)},
\end{equation}
where $l$ is the sole parameter, giving a length scale on which influence decays. 

More simple kernels include a \textit{dot-product} (or \textit{linear}) kernel
defined as
\begin{equation}
    \label{11}
        k(\mathbf{x},\:\mathbf{x}^{*})=\mathbf{x}\cdot\mathbf{x}^{*}
\end{equation}
and its more generalised form: the \textit{polynomial} kernel
\begin{equation}
    \label{12}
        k(\mathbf{x},\:\mathbf{x}^{*})=(\mathbf{x}\cdot\mathbf{x}^{*})^{\zeta},
\end{equation}
where $\zeta$ is a degree of the polynomial.

There are many other developed kernels such as \textit{Mat\'{e}rn}, \textit{periodic} and \textit{rational quadratic} whose definition conveniently found in Chapter 4 of Rasmussen's book\cite{RasmussenW06}.
% Jarv: Would be good to include a statement on what the kernels need to do. I suppose they must follow some kind of measure theory? 

In the conditioning (training) process, we may want to undertake hyperparameter optimisation, to set the observation noise $\sigma_{e}$, the kernel scale $\sigma_{o}$ and the length scale $l$ for (\ref{10}). 
One approach we implement is to maximise the logarithm of the marginal likelihood (\ref{7}): 
\begin{equation}
    \label{13}
        \begin{split}
            \log & P = -\frac{1}{2}
            \textbf{E}^{\text{T}}\cdot\left[K(\textbf{X},\:\textbf{X})+\sigma^2_{e}\mathbb{I}_{mm}\right]\cdot
            \textbf{E}
            \\& -\frac{1}{2}\log \left(\det\left[K(\textbf{X},\:\textbf{X})+\sigma^2_{e}\mathbb{I}_{mm}\right]\right)
            %\\&
            -\frac{m}{2}\log 2 \pi
        \end{split}
\end{equation}
for which its derivatives with respect to optimised hyperparameters are zero.

So far everything we have discussed is a standard Gaussian process, adapted to model a potential energy surface. 
This would require many electronic structure calculations to train the model, as we only make use of the total energy of the system (a single scalar value) per calculation. 
In the next section, we propose a derivative potential energy surface Gaussian process model, where we can additionally use the derivative of the energies (forces) which are often provided `for free' by the Hellman-Feynman theorem from electronic structure calculations. 
This significantly reduces the number of calculations required to train the model.

\subsection{Derivative model}
%\begin{center}
%\begin{figure}
%\centering
%    \includegraphics[width=0.5\textwidth]{Figures/Covariant matrix 2.png}
%    \caption{\label{fig:1} The algorithm of the model...} 
%\end{figure}
%\end{center}

To make use of force (partial derivative of energy with respect to displacement) information we need to take the derivative of the Gaussian process\cite{Solak2002,1704.00060}: kernel functions need to be differentiated with respect to the descriptors, up to the second order.

Consider the first derivative of kernel function $k(\mathbf{x},\mathbf{x}^{*})$ with respect to the descriptor $\mathbf{x}$ of $N$ atoms in Cartesian representations, $\mathbf{x}\in\mathbb{R}^{3N}$. 
The derivative can be written as a $3N\times1$ kernel matrix (Jacobian matrix as a first rank tensor)
\begin{equation}
    \label{14}
    \left[\nabla_{\mathbf{x}} k(\mathbf{x},\:\mathbf{x}^{*})\right]^{\textbf{T}}
    = \left[\frac{\partial k}{\partial x_{1}}\ \ \frac{\partial k}{\partial x_{2}}\cdots\frac{\partial k}{\partial x_{3N}} \right]^{\textbf{T}} .
\end{equation}

\begin{figure*}[t]
\centering
\includegraphics[width=0.85\textwidth]{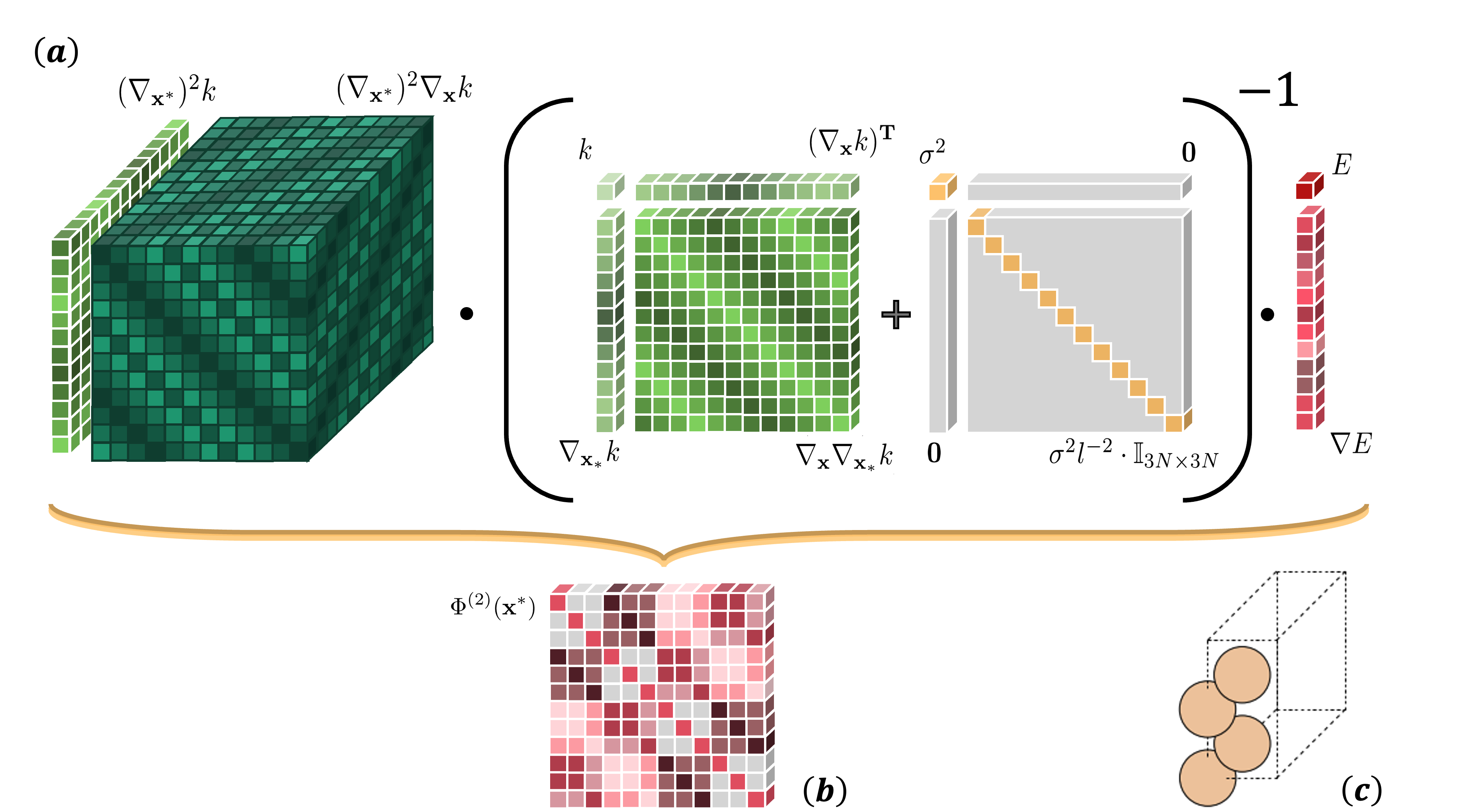}
\caption{\label{fig:1} The diagram $(1a)$ shows the tensor contraction, Eq. (\ref{20}), of the derivative GP model trained on one datapoint at $\textbf{x}$ in Cartesian coordinates. This results in figure $(1b)$, the harmonic force constant $\Phi^{(2)}$ at $\textbf{x}^{*}$, of $2\times1\times1$ Si-bulk illustrated in $(1c)$.}
\end{figure*}

In the derivative model, this matrix is used as the covariance matrix which link the force information at $\mathbf{x}$ to the potential energy at $\mathbf{x}^{*}$. 
Similarly, we can define the covariance matrix between the potential energy at $\mathbf{x}$ and forces at $\mathbf{x}^{*}$ by taking the first derivative of the kernel with respect to $\mathbf{x}^{*}$:
\begin{equation}
    \label{15}
    \left[\nabla_{\mathbf{x}^{*}} k(\mathbf{x},\:\mathbf{x}^{*})\right]^{\textbf{T}}
    = \left[\frac{\partial k}{\partial x^{*}_{1}}\ \ \frac{\partial k}{\partial x^{*}_{2}}\cdots\frac{\partial k}{\partial x^{*}_{3N}} \right]^{\textbf{T}}.
\end{equation}

Finally, we need the model marginal likelihood which contains the covariance of the forces at $\mathbf{x}$ and $\mathbf{x}^{*}$.
For this we require Hessian of the kernel with respect to those two descriptors, resulting in a $3N\times 3N$ second rank tensor covariance matrix, 
\begin{equation}
    \label{16}
    \nabla_{\mathbf{x}}\nabla_{\mathbf{x}^{*}} k
    =
    \begin{bmatrix}
            \frac{\partial^{2} k}{\partial x_{1}\partial x_{1}^{*}} & \frac{\partial^{2} k}{\partial x_{1}\partial x_{2}^{*}}& \cdots&\frac{\partial^{2} k}{\partial x_{1}\partial x_{3N}^{*}} \\
            \frac{\partial^{2} k}{\partial x_{2}\partial x_{1}^{*}} & \frac{\partial^{2} k}{\partial x_{2}\partial x_{2}^{*}}& &  \vdots\\
            \vdots &  & \ddots & \\
            \frac{\partial^{2} k}{\partial x_{3N}\partial x_{1}^{*}}& \cdots &  &  \frac{\partial^{2} k}{\partial x_{3N}\partial x_{3N}^{*}}
        \end{bmatrix}.
\end{equation}

With only one prior energy $E(\mathbf{x}_{1})$ and the corresponding forces $\nabla E(\mathbf{x}_{1})$, the marginal likelihood matrix can be reconstructed by combining (\ref{10}), (\ref{14}), (\ref{15}) and (\ref{16}). 
This marginal likelihood covariance matrix ($(1+3N)\times(1+3N)$) is 
\begin{equation}
\label{17}
\mathcal{K}(\mathbf{x}_{1},\: \mathbf{x}_{1}) = 
\begin{bmatrix}
             k & (\nabla_{\mathbf{x}} k)^{\textbf{T}}\\
             \nabla_{\mathbf{x}^{*}} k & \nabla_{\mathbf{x}}\nabla_{\mathbf{x}^{*}} k
\end{bmatrix}_{\mathbf{x},\: \mathbf{x}^{*}=\mathbf{x}_{1}}
+
\Sigma,
\end{equation}

The $(1+3N)\times(1+3N)$ $\Sigma$ observation noise matrix is 
\begin{equation}
\label{18}
\Sigma :=
\begin{bmatrix}
             \sigma^{2}_{e} & \mathbf{0}\\
             \mathbf{0} & \sigma^{2}_{f}\cdot\mathbb{I}_{3N\times 3N}
\end{bmatrix},
\end{equation}
where $\sigma_{e}$ and $\sigma_{f}$ are observation noises for prior energy and force respectively.
One choice for these noises are the energy and force convergence factor in the underlying electronic structure calculations used for training.

Therefore the potential energy prediction, analogous to (\ref{9}), is 
\begin{equation}
    \label{19}
        E(\mathbf{x}^{*}) = 
        \begin{bmatrix}
              k & (\nabla_{\mathbf{x}^{*}} k)^{\textbf{T}}
        \end{bmatrix}_{\textbf{x}^{*},\textbf{x}_{1}}
        \cdot \mathcal{K}^{-1}_{\mathbf{x}_{1},\: \mathbf{x}_{1}}
        \cdot
        \begin{bmatrix}
              E \\
              \nabla E
        \end{bmatrix}_{\mathbf{x}_{1}},
\end{equation}
where the subscripts indicate the functions evaluated at the descriptors.

Motivated by the derivative model, the differentiation can be extended to compute the second order (\textit{harmonic}) and the third order (\textit{cubic anharmonic}) force constants.
\begin{figure*}[t]
\centering
\includegraphics[width=0.85\textwidth]{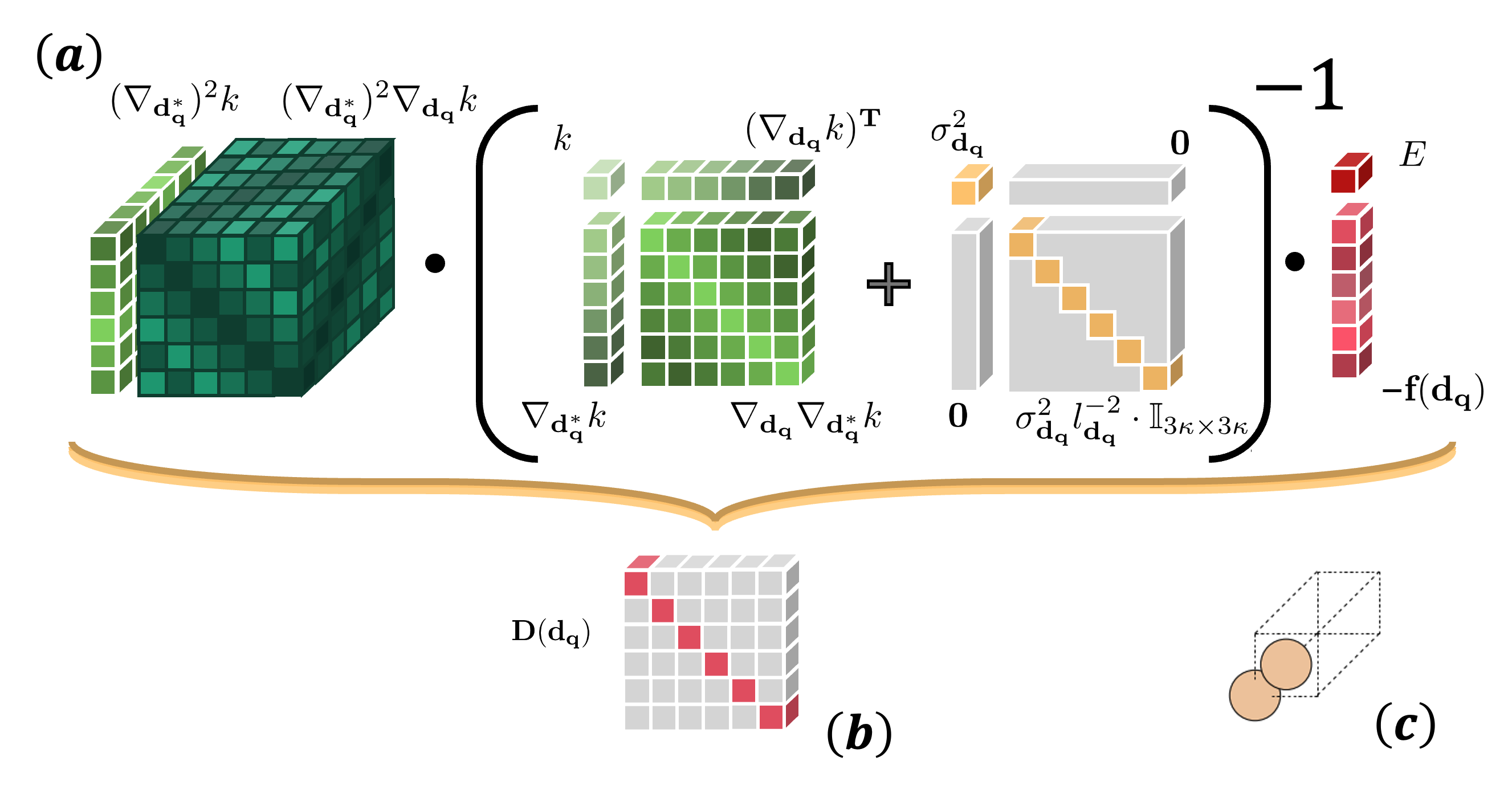}
\caption{\label{fig:2} The diagram $(2a)$ illustrates Eq. (\ref{20}) using a phonon descriptor. The results $(2b)$ is the dynamical matrix $\mathbf{D}(\mathbf{d}_{\mathbf{q}})$ instead of the harmonic force constant, which corresponds to the analysis of the primitive Si crystal $(2c)$.}
\end{figure*}
To predict the harmonic force constant, it is necessary to evaluate the second and the third order derivative with respect to those descriptors, yielding
\[(\nabla_{\mathbf{x}^{*}})^{2} k \quad \text{and}\quad (\nabla_{\mathbf{x}^{*}})^{2}  \nabla_{\mathbf{x}} k.
\]
These correspond to the $(3N\times 3N)$ rank-2 and $(3N\times 3N\times 3N)$ rank-3 tensor, which shows the covariance linking the harmonic force constant at $\mathbf{x}^{*}$ to the potential energy and to the force fields respectively at $\mathbf{x}$. We can, then, perform (\ref{19}) again with the change of those covariance matrix:
\begin{equation}
    \label{20}
    \begin{split}
        \Phi^{(2)}(\mathbf{x}^{*}) = 
        \begin{bmatrix}
              (\nabla_{\mathbf{x}^{*}})^{2} k \\(\nabla_{\mathbf{x}^{*}})^{2} \nabla_{\mathbf{x}} k
        \end{bmatrix}^{\text{T}}_{\textbf{x}^{*},\textbf{x}_{1}}
        \cdot \mathcal{K}^{-1}_{\mathbf{x}_{1},\: \mathbf{x}_{1}}
        \cdot
        \begin{bmatrix}
              E \\
              \nabla E
        \end{bmatrix}_{\mathbf{x}_{1}},
    \end{split}
\end{equation}
where $\Phi_{2}$ is the $(3N\times 3N)$ harmonic force constant containing the correlation of between all possible degree of freedoms of $\mathbf{x}^{*}$.
This Eq. (\ref{20}) can constructed as a tensor, which is then contracted, as in Figure \ref{fig:1}.

We propose a similar method for predicting the (cubic) anharmonic force constant.
The covariance tensor correlating the cubic anharmonic force constant to the potential energy and to the forces can be written as
\[(\nabla_{\mathbf{x}^{*}})^{3} k \quad \text{and} \quad (\nabla_{\mathbf{x}^{*}})^{3} \nabla_{\mathbf{x}} k\]
respectively. 
We use these rank-3 and rank-4 covariance tensors to recast (\ref{20}) as
\begin{equation}
    \label{21}
    \begin{split}
        \Phi^{(3)}(\mathbf{x}^{*}) = 
        \begin{bmatrix}
              (\nabla_{\mathbf{x}^{*}})^{3} k \\(\nabla_{\mathbf{x}^{*}})^{3} \nabla_{\mathbf{x}} k
        \end{bmatrix}^{\text{T}}_{\textbf{x}^{*},\textbf{x}_{1}}
         \cdot \mathcal{K}^{-1}_{\mathbf{x}_{1},\: \mathbf{x}_{1}}
        \cdot
        \begin{bmatrix}
              E \\
              \nabla E
        \end{bmatrix}_{\mathbf{x}_{1}}, 
    \end{split}
\end{equation}
where $\Phi_{3}$ is the cubic anharmonic force constant tensor with the size of $(3N\times 3N\times 3N)$. 

With these methods of directly evaluating the force constants from our fitted Gaussian Process, we have all the necessary components for calculating the lattice dynamics of a material, including anharmonicity.

\begin{figure*}[t]\centering
\captionsetup{justification=centering}
    \begin{tabular}{c}
        \includegraphics[width=0.45\linewidth,height=\textheight,keepaspectratio]{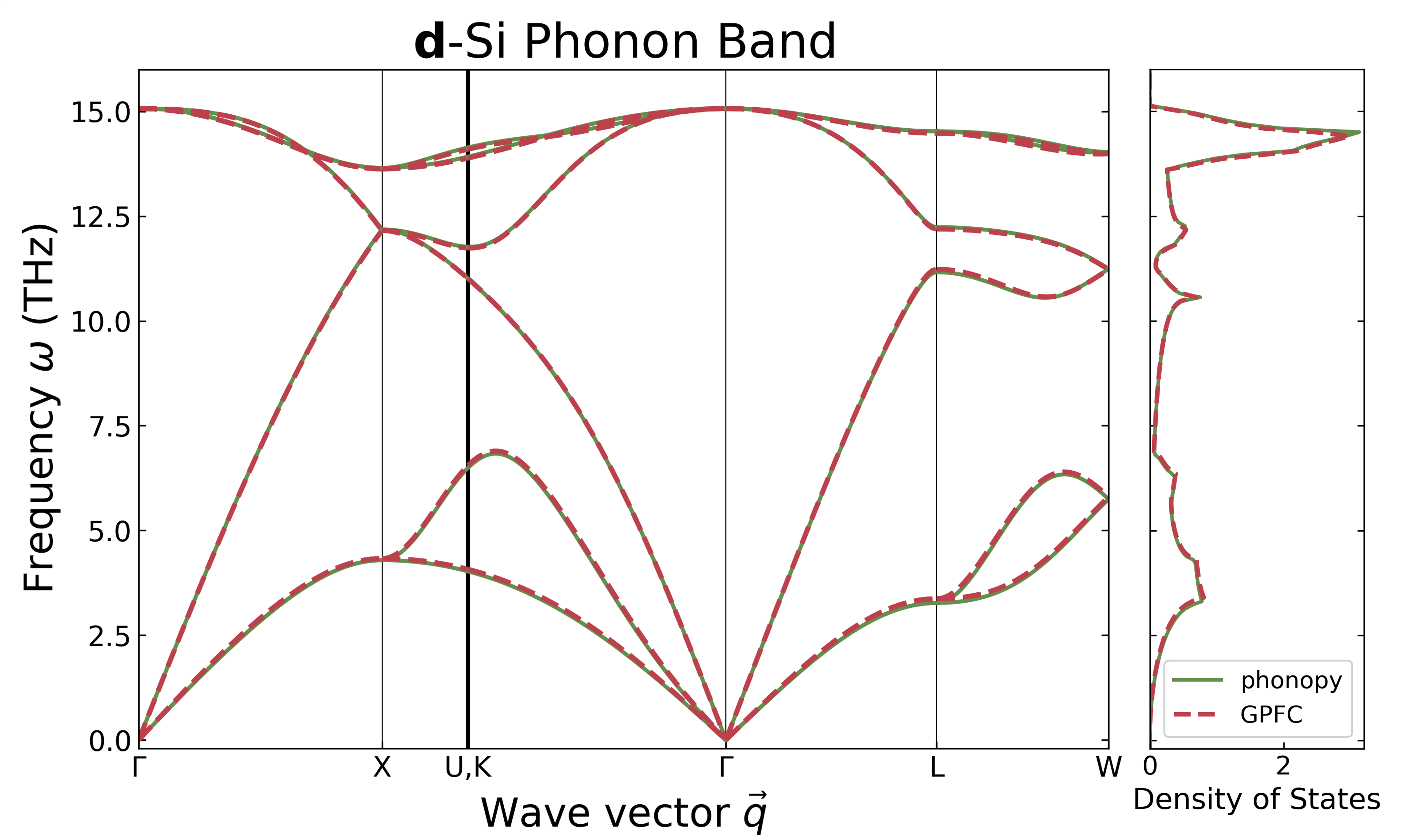}  \\
        (a)\\[6pt]
    \end{tabular}

    \begin{tabular}{cc}
        \includegraphics[width=0.45\linewidth,height=\textheight,keepaspectratio]{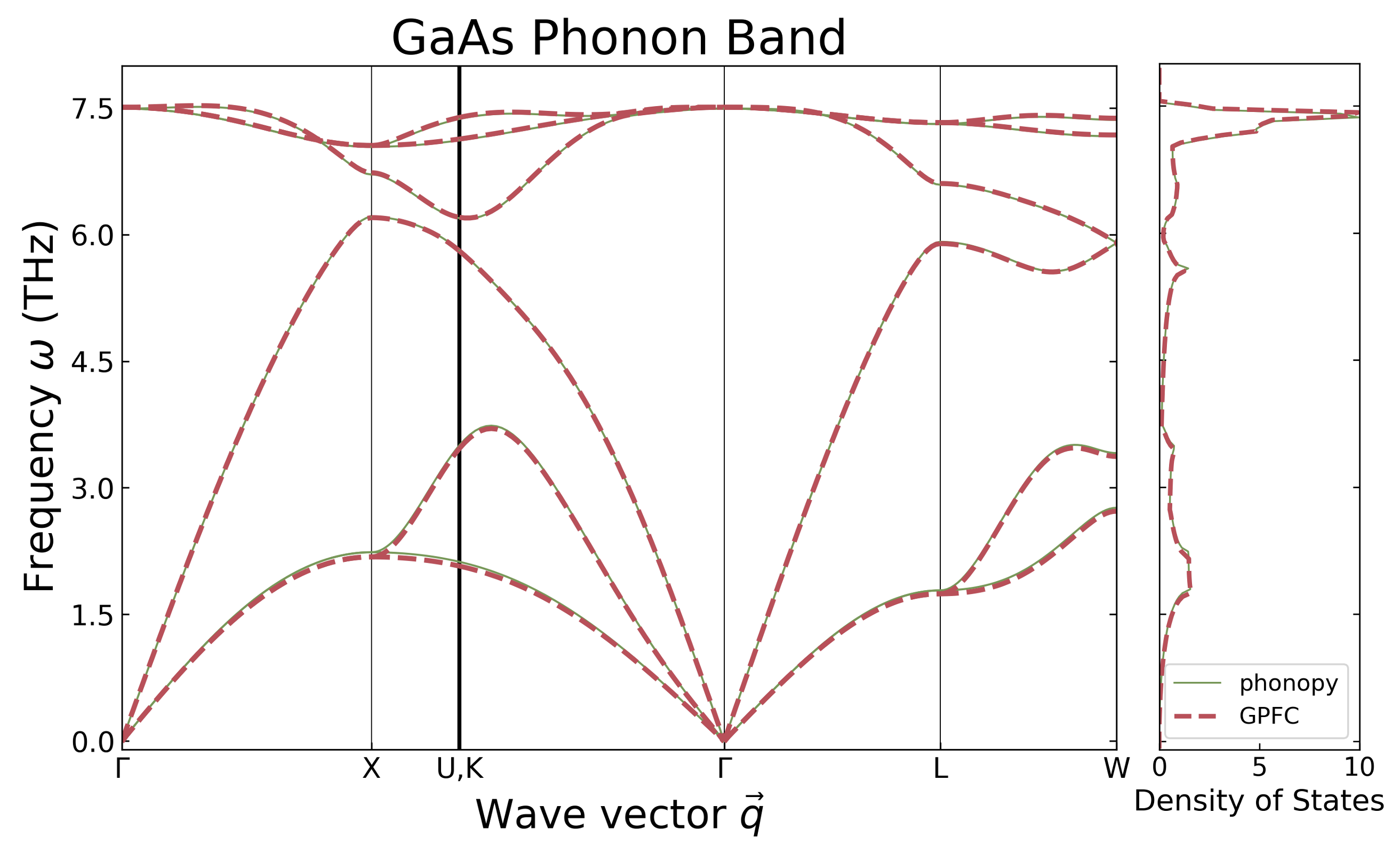}&
        \includegraphics[width=0.45\linewidth,height=\textheight,keepaspectratio]{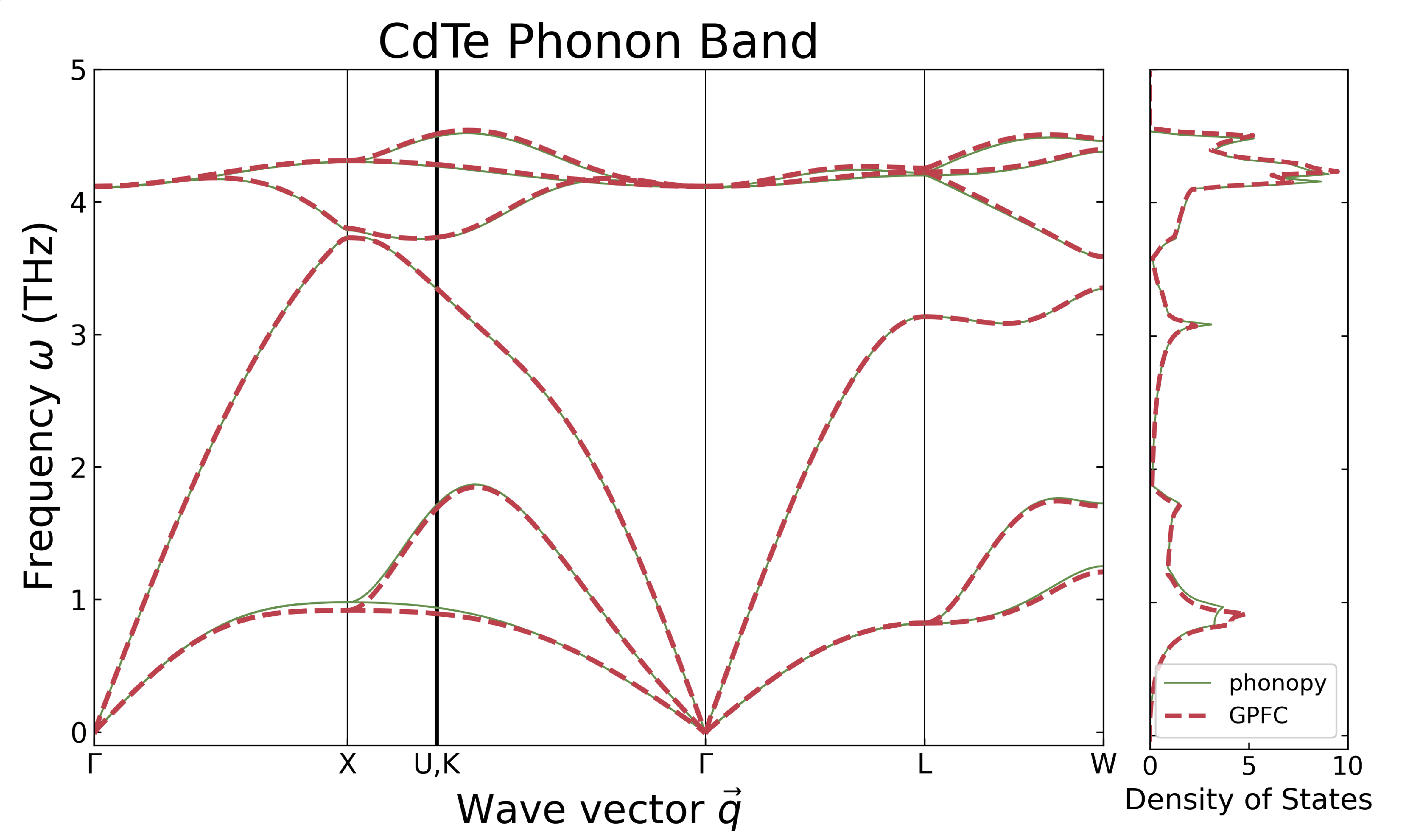}\\
        (b) & (c)\\[6pt]
    \end{tabular}

    \begin{tabular}{cc}
        \includegraphics[width=0.45\linewidth,height=\textheight,keepaspectratio]{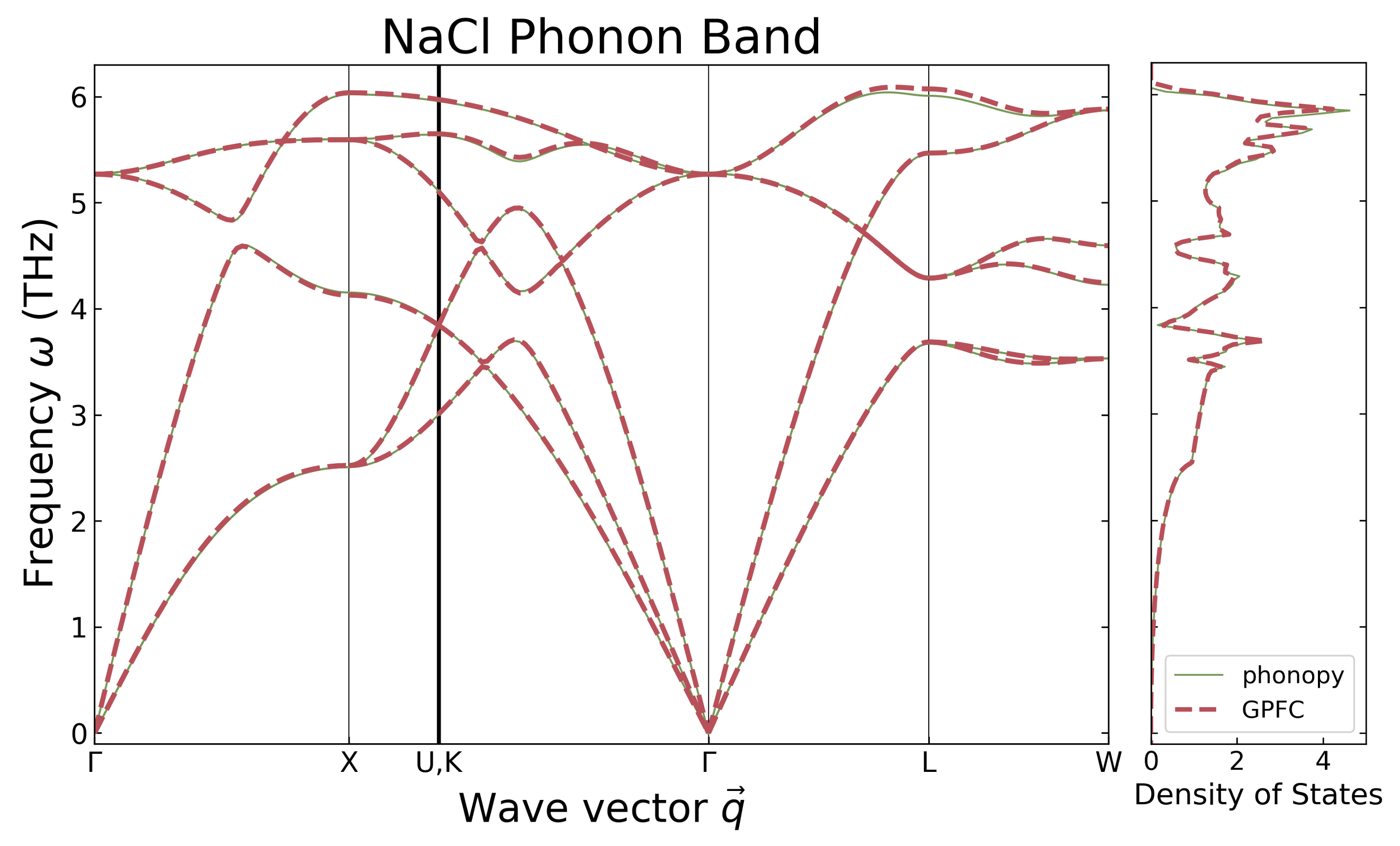} &
        \includegraphics[width=0.45\linewidth,height=\textheight,keepaspectratio]{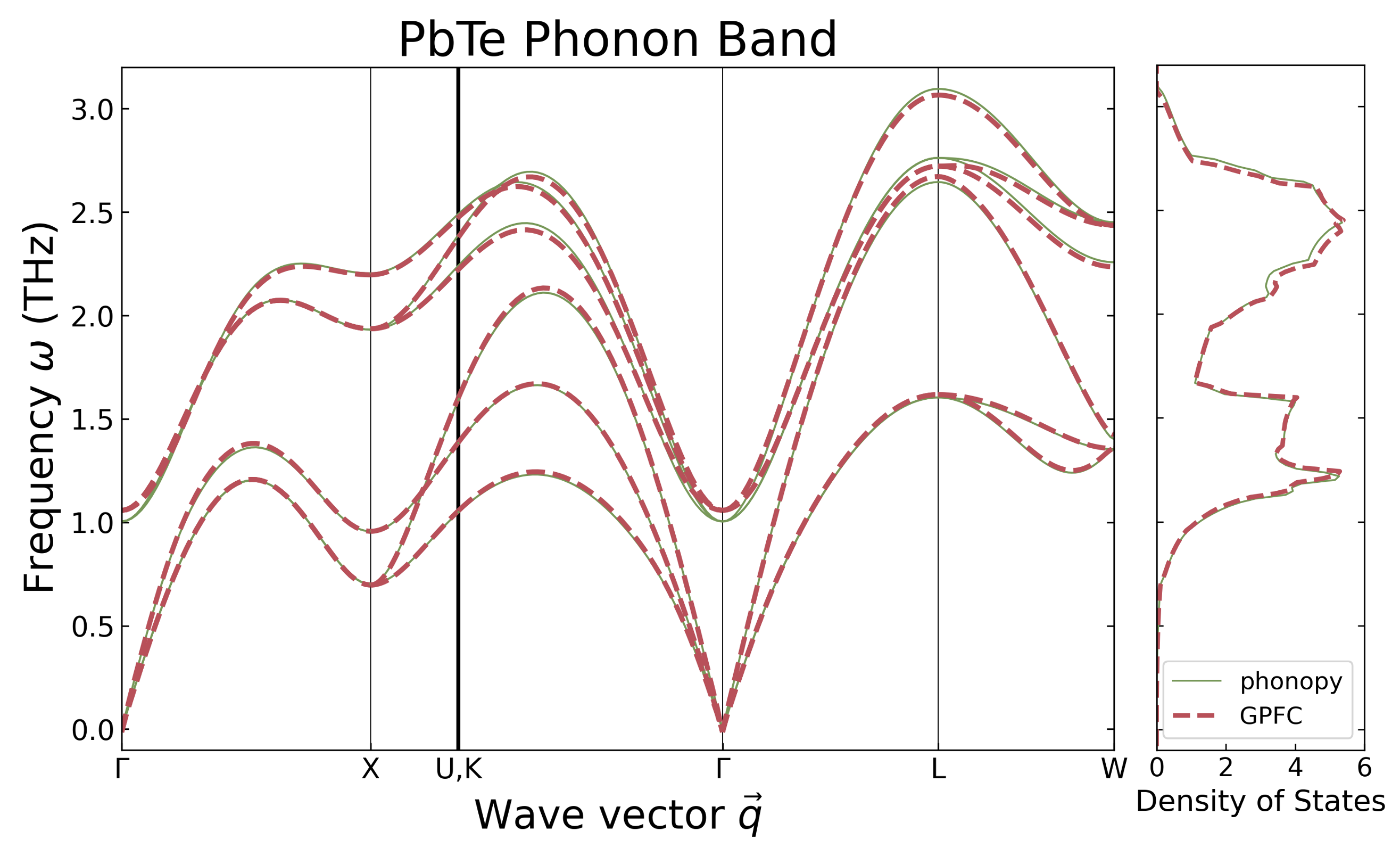} \\
        (d) & (e)\\[6pt]
    \end{tabular}

    \caption
    { \label{fig:3}
    Figure (a), (b), (c), (d) and (e) show the phonon band structures and density of states of d-\textbf{Si} (FCC), \textbf{GaAs}, \textbf{CdTe}, \textbf{NaCl} and \textbf{PbTe} respectively. Solid green lines are calculated from the harmonic force constants evaluated by \texttt{Phonopy} (with finite displacement methods), while the dashed red lines are calculated from the harmonic Gaussian process force constants.
    }
\end{figure*}

\subsection{Phonon coordinate representations}

Instead of using the direct Cartesian representations, it seemed to be a good idea to try and directly use the phonon coordinates (normal modes) with the Gaussian process. 
These normal modes form a natural basis for the motion of a material or molecule, and naturally encode the symmetries of the material or molecule. 
Due to this, they are also the starting point for higher order physical models of lattice dynamics, such as many-body perturbation-theory which couples the normal modes through the anharmonicity. 

Our hope was that this representation would increase the efficiency of the machine learning method (i.e. greater accuracy for fewer evaluations), without having to build any symmetries or equivariances directly into our methods. 
The normal modes are provided by diagonalising the mass-weighted second-order force-constant matrix (the dynamic matrix).  

\begin{figure*}[t]\centering
\captionsetup{justification=centering}
    \begin{tabular}{cc}
        \includegraphics[width=0.45\linewidth,height=\textheight,keepaspectratio]{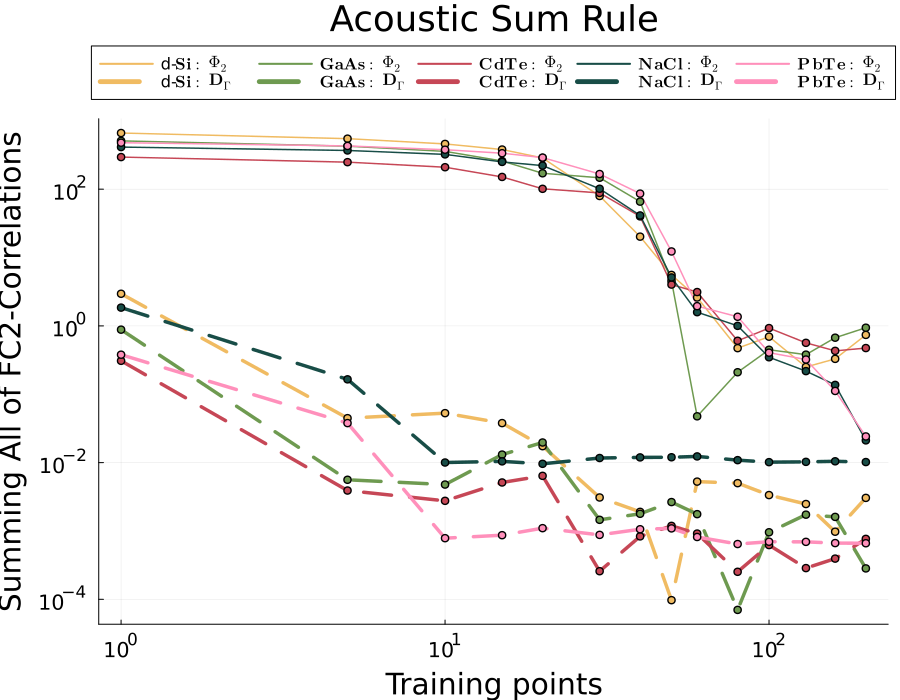} &
        \includegraphics[width=0.45\linewidth,height=\textheight,keepaspectratio]{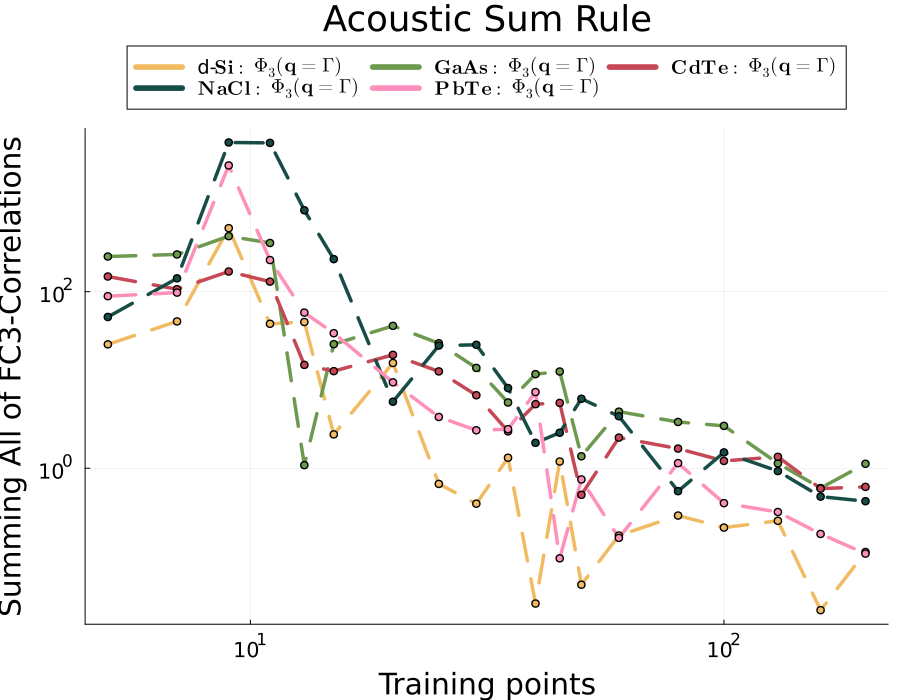} \\
        (a) &  (b)\\[6pt]
    \end{tabular}
    \caption
    { \label{fig:4}
    Panel (a) and (b) show the FC2 and FC3 learning curve via the acoustic sum rule. 
    Yellow, red, green, blue and pink colours represent d-\textbf{Si} (FCC), \textbf{GaAs}, \textbf{CdTe}, \textbf{NaCl} and \textbf{PbTe}.
    The solid and dashed lines indicate the force constant in Cartesian and Phonon basis (evaluated at gamma: $\Gamma$ point).
    }
\end{figure*}

To build phonon descriptors, we start by considering this ``\textit{Dynamical Matrix}'' \cite{phonopy, Togo2015-qf, Phonopy_toolkit} at wave vector $\textbf{q}$:
\begin{equation}
    \label{22}
    \begin{split}
        D_{Ai,Bj}(\textbf{q})= 
        \frac{1}{\sqrt{m_{A}m'_{B}}}
        \sum_{l'}\Phi^{(2)}_{0Ai,l'Bj}
        \textbf{e}^{i\textbf{q}\cdot(\textbf{R}^{o}_{l'B}-\textbf{R}^{o}_{0A})}
    \end{split}
\end{equation}
for component $i$ of atom $A$ and component $j$ of $B$ with mass $m_{A}$ and $m'_{B}$ respectively.
The summation is over all possible primitive cells $l'$ in the supercell where $0$ indicates the reference primitive cell.
Here $\textbf{R}^{o}$ is an atomic displacement from the equilibrium.

Our phonon descriptors are derived from a simple change of basis, using these new eignevectors of the dynamics matrix, 
\begin{equation}
    \label{23}
    \begin{split}
        d_{Bj}(\textbf{q})= 
        \sqrt{m'_{B}}
        \sum_{l'}\textbf{x}_{l'Bj}
        \textbf{e}^{i\textbf{q}\cdot(\textbf{R}^{o}_{l'B}-\textbf{R}^{o}_{0A})},
    \end{split}
\end{equation}
where $\textbf{x}$ are the original Cartesian descriptors \cite{Kohei_spgrep}.

The system symmetry is directly imposed to the model: the model does not have to learn the symmetry by itself.
Additionally, the model descriptors are collapsed from an $l\times l\times l$ supercell to a $\kappa$-atom primitive cell size; i.e., $N = l^{3}\kappa$ atoms are in the atomic environment.
The corresponding tensor contraction Equation (\ref{20}) can be reconstructed as shown in Figure \ref{fig:2}.

For $m$ training points, the learning costs of $\mathcal{O}(3^{3}m^{3}N^{3})$ and the prediction cost $\mathcal{O}(3mN)$ \cite{RasmussenW06, David} are reduced to $\mathcal{O}(3^{3}m^{3}\kappa^{3})$ and $\mathcal{O}(3m\kappa)$ respectively.

The predicted second order force constant (as a yield from Equation (\ref{20})) will be in phonon coordinates as a dynamical matrix at wave-vector $\mathbf{q}$. We can then recover the second order force constant in Cartesian coordinates by using the inverse discrete Fourier transform of the set of $N$ dynamical matrices, Equation (\ref{22}), as following
\begin{equation}
    \label{24}
    \begin{split}
        \Phi^{(2)}_{0Ai,l'Bj}= 
        \frac{\sqrt{m_{A}m'_{B}}}{N}
        \sum_{\textbf{q}}D_{Ai,Bj}(\textbf{q})
        \textbf{e}^{-i\textbf{q}\cdot(\textbf{R}^{o}_{l'B}-\textbf{R}^{o}_{0A})} .
    \end{split}
\end{equation}

Similarly, the third order force constant predicted in phonon coordinates can be transform back to one in Cartesian coordinates with the similar expression, 
\begin{equation}
    \label{25}
    \begin{split}
        \Phi^{(3)}_{0Ai,l'Bj,l''Ck}=&
        \frac{\sqrt{m_{A}m'_{B}m''_{C}}}{N}
        \sum_{\textbf{q}',\textbf{q}''}\Phi^{(3)}_{Ai,Bj,Ck}(\textbf{q}',\textbf{q}'')\\
        &\textbf{e}^{-i\textbf{q}'\cdot(\textbf{R}^{o}_{l'B}-\textbf{R}^{o}_{0A})}
        \times\textbf{e}^{-i\textbf{q}''\cdot(\textbf{R}^{o}_{l''C}-\textbf{R}^{o}_{0A})}\\
        &\textbf{e}^{-i(\textbf{q}+\textbf{q}'+\textbf{q}'')\cdot\textbf{R}^{o}_{0A}}\Delta(\textbf{q}+\textbf{q}'+\textbf{q}'') ,
    \end{split}
\end{equation}
where these wave vectors have to conserve the lattice momentum. This means they are confined by
\begin{equation}
\Delta(\textbf{q}+\textbf{q}'+\textbf{q}'')=
    \label{26}
    \begin{cases}
        1 & \text{$\textbf{q}+\textbf{q}'+\textbf{q}''$ is reciprocal}\\
          & \quad \text{lattice vector}\\
        0 & \text{Otherwise}.
    \end{cases}
\end{equation}

Therefore, the three phonon descriptors ($d_{Ai}(\textbf{q})$, $d_{Bj}(\textbf{q}')$ and $d_{Ck}(\textbf{q}'')$ ) used in our prediction have to satisfy this constraint.

\section{Dataset preparation and training}

In the development of our methods, and to compare to standard approaches, we consider a set of representative solid state materials. 
We compare our method (which we term GPFC, Gaussian Process Force-Constants) with a traditional finite displacement method (as implemented in \textsc{Phonopy} \cite{phonopy,Phonopy_toolkit}) and a third-order many-body perturbation theory approach (in \textsc{Phono3py} \cite{Togo2015-qf,Phonopy_toolkit}).%; and with a cluster-expansion approach (in \texttt{HiPhive}). 
For processing of our predicted force-constants we use \textsc{Phonopy} and \textsc{Phono3py}. 
The details of the training datasets are provided below.

\textsc{phono(3)py}-dataset: we use \textsc{phono(3)py}-package to generate displaced structures of $2\times2\times2$ supercells following finite different methods (FDM), and then evaluate the energy and force in a density functional theory calculation with \textsc{VASP}, with plane-wave cutoff energy (\textsc{ENMAX} $= 800\:eV$), SCF energy convergence (\texttt{EDIFF} $= 10^{-8}\:eV$), and k-point (\textsc{kpts} $=[6,\:6,\:6]$).
After \textsc{VASP} processes, we then use \textsc{VASP} results with \textsc{Phonopy}-package to generate force constants with space group symmetry operation.

% skipping HiPhive data for now, until we generate the figures
%\blue{\textsc{HiPhive}-dataset: we use \textsc{HiPhive} data generation function to obtain rattled structures with normal distributed $0.01\:{\AA}$ and perform similar density functional theory calculations of energy and force in \textsc{VASP}.
%The plane-wave energy cutoff and convergence parameters are as above.
%We set k-point mesh of \textsc{kpts} $=[6,\:6,\:6]$ and $=[2,\:2,\:2]$ for $2\times2\times2$ supercells (lower cutoff distance) and $5\times5\times5$ supercells (higher cutoff distance), respectively}

%The force data from the calculations are used to train the machine learning model in \texttt{HiPhive} to parameterise force constants along with space group symmetry through translational and rotational sum-rule relation.

\textsc{GPFC}-dataset: we use \textsc{ASE} to generate rattled ($2\times2\times2$ supercell) structures of the atomic environments with normal distribution $0.01\:{\AA}$, and then perform similar density functional theory calculations of energy and force in \textsc{VASP}.
The plane-wave energy cutoff, k-point mesh, convergence parameters are as above.
Subsequently, the dataset of total energies and forces are used to train the derivative GP model following Eq (\ref{20}) without imposing any symmetry.

%\begin{figure}[t]
%\centering
%\includegraphics[width=0.45\textwidth]{Figures/FC2_figures/FC2_sumrule_log.png}
%\caption{\label{fig:4} This figure shows the FC2 learning curve via the acoustic sum rule. Yellow, red, green, blue and pink colours represent d-\textbf{Si} (FCC), \textbf{NaCl}, \textbf{PbTe}, \textbf{GaAs} and \textbf{CdTe}, %while the solid and dashed lines indicate the force constant in Cartesian and Phonon basis respectively}
%\end{figure}

Though the GPFC hyperparameters (kernel scale ($\sigma_{o}$), a length scale ($l$), the observation noise of energy ($\sigma_{e}$) and force ($\sigma_{f}$)) can be optimised by maximising the logarithm of the marginal likelihood as in Eq. (\ref{13}), we set them to reasonable constants. 
$\sigma_{o}$ and $l$ are $1$ and $0.4$ respectively for the $0.01\:-\:0.05\:\AA$ normal distributed rattles of the relaxed structure \cite{GarijodelRo2019}.
The observation noises $\sigma_{e}$ and $\sigma_{f}$ we taje as $10e-8$, the same as our SCF energy convergence (\texttt{EDIFF}) in our \texttt{VASP} calculations. 

\begin{figure*}[t]
\captionsetup{justification=centering}
    \begin{tabular}{cc}
        \includegraphics[width=0.5\linewidth,height=\textheight,keepaspectratio]{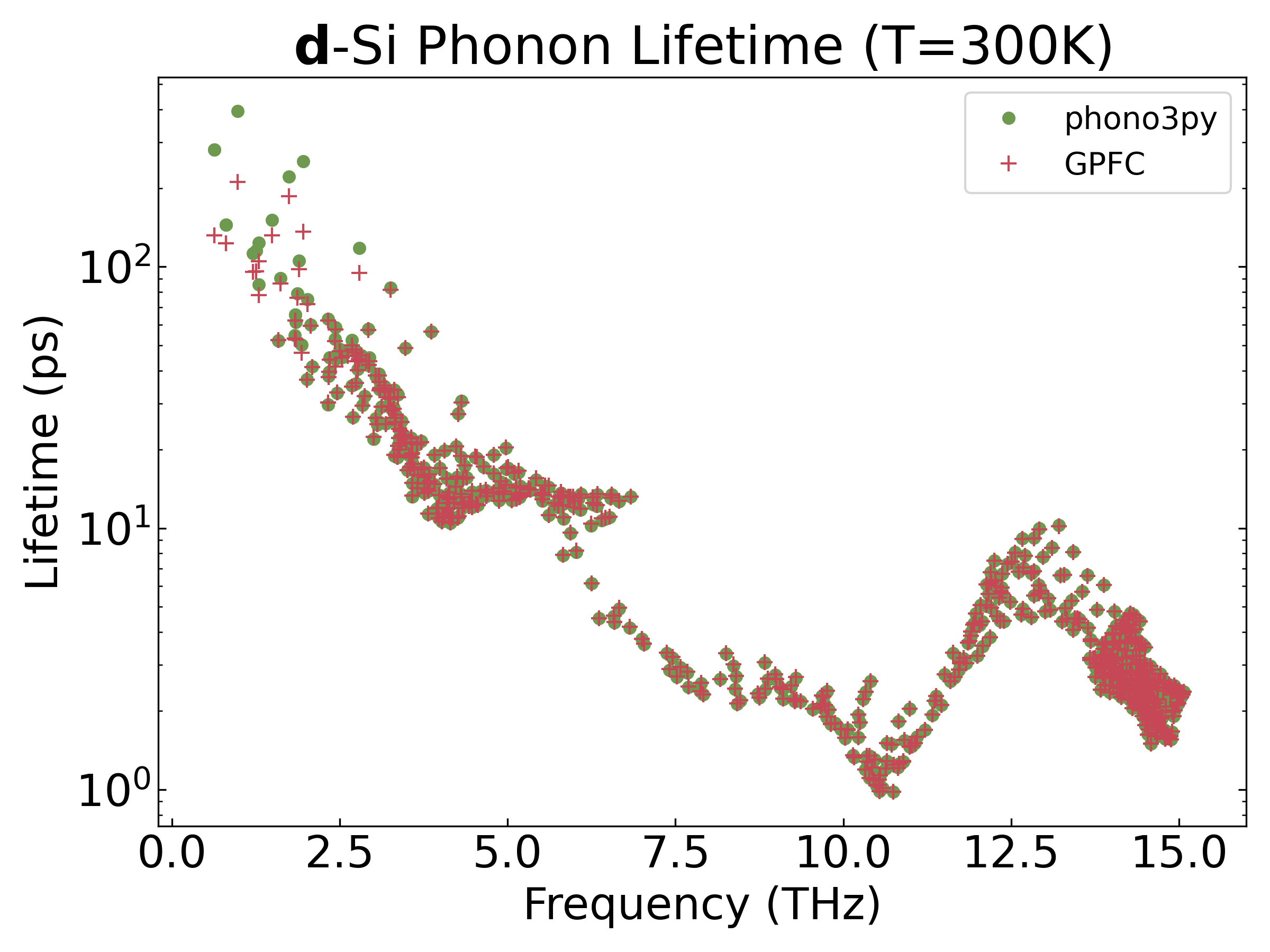}&
        \includegraphics[width=0.5\linewidth,height=\textheight,keepaspectratio]{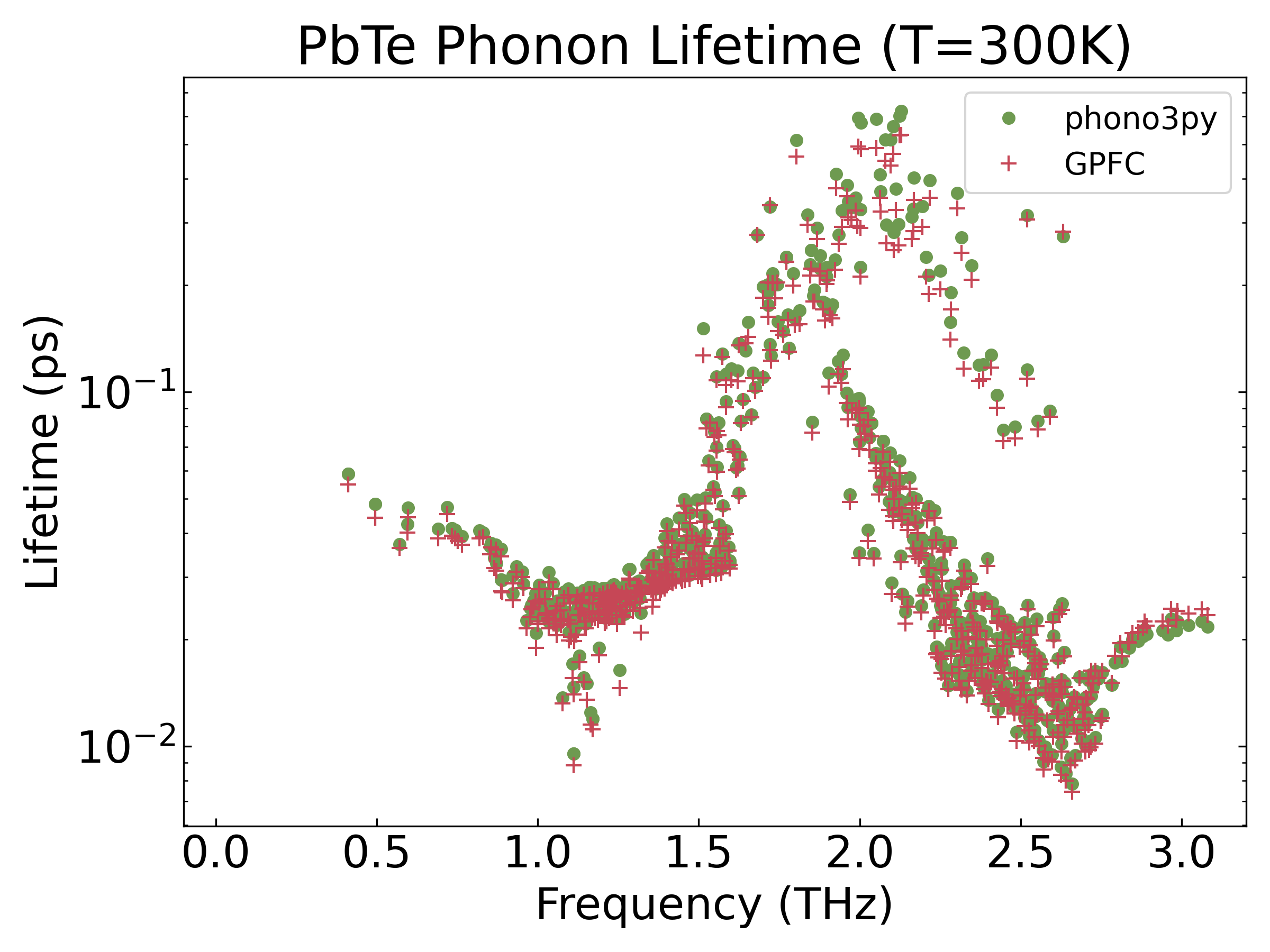}\\
        (a) & (b)\\[6pt]
    \end{tabular}
    
    \begin{tabular}{c}
        \includegraphics[width=0.45\linewidth,height=\textheight,keepaspectratio]{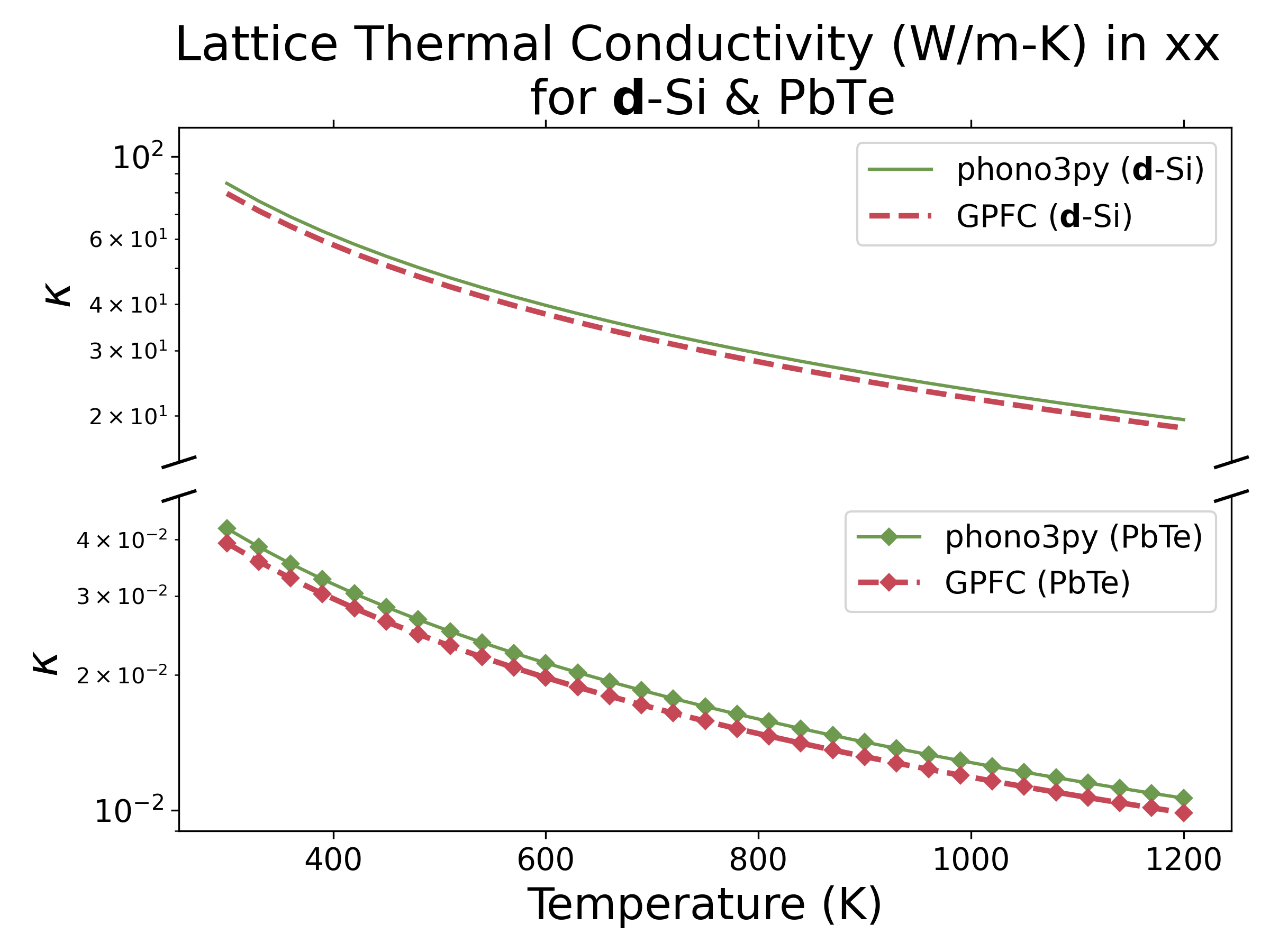}\\
        (c) \\[6pt]
    \end{tabular}

    \caption
    { \label{fig:5}
    These figures illustrate the predicted lifetime of phonons at $300$~K and the lattice thermal conductivity of d-\textbf{Si} and \textbf{PbTe} from $300\:K$ to $1200\:K$. Green point clouds and lines are calculated from the cubic anharmonic force constants evaluated by \texttt{Phono3py} (with finite displacement methods), while Red point clouds and lines are calculated from the cubic anharmonic Gaussian process force constants.
    }
\end{figure*}

\section{Results and discussion}

To compare the harmonic (second order) force constants between our model and the standard finite-displacement method (FDM) as implemented in \texttt{Phonopy}, we examine the phonon band diagram and its density of state of a diamond(d)-\textbf{Si} (FCC), \textbf{GaAs}, \textbf{CdTe}, \textbf{NaCl} and \textbf{PbTe}, which are illustrated in Figure \ref{fig:3}.
The root mean square errors of the phonon band structures are around $1-4\%$.
% Jarv - what does RMSD mean here? Integrating across k-space? But the bands are just cuts through the BZ. 
Wasserstein's distance (earth mover distance, EMD) is used to measure the dissimilarity of two phonon density of states between different methods over a frequency $\omega$ region: we calculate these as $0.043$, $0.037$, $0.021$, $0.071$ and $0.022$ $THz\cdot eV^{-1}\AA^{-3}$ respectively. 
%(Need a paper from refer to the acceptable EMD for pDOS)

The model learning curves are shown via calculating the degree of acoustic (or translational) sum-rule violation. 
For harmonic force-constants, this should be 
\begin{equation}
    \label{27}
    \begin{split}
        \sum_{B}\Phi^{(2)}_{Ai,Bj} = 0,
    \end{split}
\end{equation}
and for third order anharmonic force-constants this would be 
\begin{equation}
    \label{28}
    \begin{split}
        \sum_{C}\Phi^{(3)}_{Ai,Bj,Ck} = 0, 
    \end{split}
\end{equation}
with components $i$,$j$,$k$, of atoms $A$,$B$,$C$ respectively.

In Figure \ref{fig:4} (a), we can see the sum-rules in the Cartesian basis for all those three materials start to converge to zero with $\approx 48$ data-points.
This number corresponds to the degree of freedoms of the atomic descriptor in Cartesian coordinates, which are $48$ for the $2\times2\times2$ supercell of a 2-atomic primitive cell.
Although those five materials are different, their learning curves converge with the same number of training points.

Meanwhile in phonon coordinates, they converge to zero with $\approx 6$ datapoints. 
The number datapoints required similarly correspond to the degree of freedoms of the phonon atomic descriptor (now in the primitive unit-cell).
Phonon coordinate FC2 (dynamical matrix) and FC3 are evaluated at $\Gamma$-point.
In other \textbf{q}-points, the evaluations are limited by current automatic differentiation package.
With this limitation, we cannot recover Cartesian FC2 and FC3 with using Eq (\ref{24}) and Eq (\ref{25}).
%\begin{table}[]
%\begin{tabular}{ |c|c|c|c|c|}
% \hline d-\textbf{Si}   & \textbf{NaCl}  & \textbf{PbTe}  & \textbf{CdTe}   & \textbf{GaAS}   \\  \hline
% \hline 0.04342 & 0.07107 & 0.02237 & 0.03711 & 0.02071 \\ \hline
%\end{tabular}
%\caption{\label{table:1} This table shows Wasserstein's distance (earth mover's distance: EMD) used to compare phonon density of states (pDOS) of those materials among different methods, comparing %pDOSs from \texttt{GPFC.jl}}
%\end{table}
%\begin{figure}[t]
%\centering
%\includegraphics[width=0.45\textwidth]{Figures/FC3_figures/FC3_sumrule_log.png}
%\caption{\label{fig:5} This figure shows the FC3 learning curve via the acoustic sum rule. Yellow, red, green, blue and pink lines correspond to the force constant of d-\textbf{Si} (FCC), \textbf{NaCl}, \textbf{PbTe}, %\textbf{GaAs} and \textbf{CdTe} in phonon coordinates.}
%\end{figure}

To compare the cubic anharmonic (or third order) force constants among the methods, we consider the lifetime of the phonons and the lattice thermal conductivity of d-\textbf{Si}, \textbf{NaCl}, \textbf{PbTe}, \textbf{GaAs} and \textbf{CdTe} with finite temperature ($300-1200$~K).
They are calculated by using a third-order many-body perturbation theory approach with different cubic anharmonic force constants, one from FDM in \texttt{Phono3py} and another from kernel regression in our \texttt{GPFC}.
The cubic anharmonic Gaussian process force constant is predicted by using Eq. (\ref{23}).
We train the derivative GP model with the same training set as using in the prediction of the harmonic force constants.

Based on the convergence of the acoustic sum-rule (Figure \ref{fig:4}), 
accurate prediction of the cubic anharmonic force constants with a Cartesian basis requires up to $400$ energies and force calculations. 
Meanwhile, in the phonon basis, starting with a dynamic-matrix (harmonic force constants), only $50$ energy and forces are required. 

% Jarv: Phonon Basis here, or Cartesian?
The predicted lifetime of phonons of d-\textbf{Si} and \textbf{PbTe} at $300$~K are shown in panel \ref{fig:5} (a) and \ref{fig:5} (b), while their predicted lattice thermal conductivity illustrated in Figure \ref{fig:5} (c). 
Again, the Earth Mover Distance is used to quantify dissimilarity of the projected lifetime density of state, for each finite $30$~K step from $300$~K to $1200$~K.
The mean EMD for d-\textbf{Si} is $3.9\times10^{-1}\:ps$, while $8.6\times10^{-3}\:ps$ is for \textbf{PbTe}.
%EMD $0.0312$ (NaCl), $0.0299$ (GaAs), $0.0807$ (CdTe)
Mean absolute errors of the lattice thermal conductivity calculated by using the third order GPFC are $2.03$ ($\sim 2\%$ error) and $1.4\times10^{-3}$ ($\sim 7\%$ error)  $W\cdot m^{-1}K^{-1}$, respectively.
Moreover, the lifetime of phonons and the lattice thermal conductivity of \textbf{NaCl}, \textbf{GaAs} and \textbf{CdTe} are calculated based on our GPFCs with the errors ranging from $2\%$ to $7\%$.
The accuracy of the third order GPFC prediction for \textbf{PbTe} is low comparing among five materials because \textbf{PbTe} exhibits strong anharmonic behaviour leading to phonon-phonon interaction which can reduce its phonon lifetime.
%The effect along with Gaussian processes being noisy could result in the lowest prediction accuracy comparing among these five materials.
%MAE  $0.1029$ (NaCl), $0.4070$ (GaAs), $0.03739$ (CdTe)

A key finding of our experiments with our GPFC approach is that we seem to require a number of training points equal to the degree of freedom of the descriptors. In the phonon basis this is the number of phonon bands which you are calculating. 
Potentially the Coulomb matrix \cite{Rupp2012}, or similar species-aware descriptor, could be used to share information about similar chemical species in the unit cell.

In our anharmonic (to cubic) GPFC experiments, we find we require approximately $8$ times the number of data points required for a harmonic GPFC. 

Therefore, in our limited experiments the calculation of anharmonic force constants is linear in the number of elements in the unit cell.

\section{Conclusion}

We develop a method to model lattice dynamics, by fitting a \textit{derivative} Gaussian Process force-constant (GPFC) model. 
These derivatives provide the correlation among energy and its derivatives, i.e. force, second-order force constants (FC2), and FC3. 
We experiment with this model on the harmonic and anharmonic (cubic) force constants of five materials (d-\textbf{Si}, \textbf{NaCl}, \textbf{PbTe}, \textbf{GaAs} and \textbf{CdTe}). 
Accurate predictions of the harmonic force constants require the same number of energy and force evaluations (here density functional theory calculations) as the number of degrees of freedom in our descriptor basis. 
For the most compact phonon descriptor, this means the same number as the phonon bands to be predicted. 
To predict the third-order force constants, we seem to require 8 times the data than required to fit the harmonic force constants. 

Our models seem to offer linear scaling of prediction of anharmonic force constants. 
We are technically limited in being able to apply our phonon descriptor to anharmonic force constants, due to the necessity of dealing with the derivative of a complex number at positions away from gamma in the Brillouin-Zone. 

An extension of this work would be to leverage the development of atomistic machine-learning force-field descriptors, i.e. radial basis and spherical Harmonic basis functions as in \textsc{GAP} \cite{Bartk2010,Bartk2013} or \textsc{ACE} \cite{Drautz2019, Witt2023-wg}. 
These model descriptors are equivariant under the rotation in a three dimensional space, and may offer some of the benefits of the phonon descriptors we have developed in this work. 
%Instead of constructing $O(3)$-irreducible representations, we can use the $O(3)$-equivariant basis function available in \texttt{Julia}: namely, \texttt{ACEpotential.jl} \cite{Witt2023-wg}.

As an alternative approach, one could directly train (or fine tune) a full machine-learning force-field, and then use this as a surrogate model for the standard finite displacement phonon workflow. In a future study we will compare the data efficiency of this approach.

%Future work: (not for this paper).
%\subsection{Spherical Harmonic Representations}
%We aim to use Radial basis and Spherical Harmonic basis to represent the model descriptor as in \texttt{GAP} \cite{Bartk2010,Bartk2013}, \texttt{ACE} \cite{Drautz2019, %Witt2023-wg} and \texttt{MACE} \cite{MACE,MACE2}.
%These model descriptors are equivariant under the rotation in a three dimensional space. 
%Instead of constructing $O(3)$-irreducible representations, we can use the $O(3)$-equivariant basis function available in \texttt{ACEpotential.jl} \cite{Witt2023-wg}.

\section{Author contribution}

K.K.: 
Formal Analysis (lead); 
Investigation (lead); 
Methodology (equal);
Software (lead);
Writing – original draft (equal); 
Writing – review and editing (equal).

J.M.F.: 
Conceptualization (lead); 
Methodology (equal);
Writing – original draft (equal); 
Writing – review and editing (equal).

\section{Acknowledgement}
J.M.F. is supported by a Royal Society University Research Fellowship
(URF-R1-191292). 
K.K. is supported by a Thai scholarship, Development and Promotion of Science and Technology project. 
Julia\cite{Julia} codes implementing these calculations are available as a repository on
GitHub\cite{GitHub}. 
This work made use of the Imperial College Research Computing Service \cite{HPC}.
Via our membership of the UK's HEC Materials Chemistry Consortium, which is funded by EPSRC (EP/R029431 and EP/X035859), this work used the \textsc{ARCHER2} UK National Supercomputing Service (http://www.archer2.ac.uk).

\onecolumngrid % need to use onecolumngrid; as APS disables onecolumn
\newpage
\appendix 

\section{Force constants from a Gaussian Process}

\bibliography{GaussianProcessPotentialEnergySurface}

\end{document}